\documentclass[a4paper,12pt]{article}

\usepackage{amsmath, amssymb, amscd, amsthm}
\usepackage{mathrsfs}
\usepackage[font=small,skip=0pt]{caption}

\usepackage{bm}
\usepackage{csquotes, color}
\usepackage{colortbl}
\usepackage{tocloft}
\usepackage[numbers]{natbib}
\usepackage{verbatim}
\usepackage[toc,page]{appendix}
\usepackage{caption}
\usepackage{textcomp}
\usepackage[inline]{enumitem}
\usepackage{amssymb}
\usepackage{tcolorbox}
\usepackage{cancel}

\usepackage{authblk}

\usepackage{algorithm}
\usepackage{algorithmic}
\usepackage{scrextend}
\addtokomafont{labelinglabel}{\sffamily}

\usepackage{gnuplot-lua-tikz}
\usetikzlibrary{arrows.meta}
\usetikzlibrary{positioning}
\usetikzlibrary{shadows}
\usetikzlibrary{calc}
\usetikzlibrary{shapes.multipart}
\usetikzlibrary{patterns}

\newtheorem{definition}{Definition}
\newtheorem*{definition*}{Definition}
\newtheorem{remark}{Remark}
\newtheorem*{conditional*}{Conditional}
\newtheorem{conditional}{Conditional}
\newtheorem*{remark*}{Remark}

\tikzset{Myarrow/.style={very thin, arrows=-Latex}}

\definecolor{Gray}{gray}{0.85}
\definecolor{LightCyan}{rgb}{0.88,1,1}

\newcolumntype{a}{>{\columncolor{Gray}}c}
\newcolumntype{b}{>{\columncolor{Gray}}r}
\newcolumntype{d}{>{\columncolor{white}}c}

\newcommand{\cF}{{\cal F}}

\newcommand{\cH}{{\cal H}}

\newcommand{\cP}{{\cal P}}

\newcommand{\scrL}{{\mathscr L}} 
 
\newcommand{\scrV}{{\mathscr V}} 

\usepackage{hyperref}
\hypersetup{
    colorlinks=true,
    citecolor=black,
    linkcolor=black,
    filecolor=cyan,      
    urlcolor=blue,
}
 
\title{Growth of Random Trees by Leaf Attachment 
}
\author{ Nomvelo Karabo Sibisi}
\affil{University of Cape Town \\ {\small {\tt sbsnom005@myuct.ac.za}}}
\date{October 2020}

\begin{document}
\maketitle
\thispagestyle{empty}

\begin{abstract}
\noindent
We study the  growth of a  time-ordered  rooted tree by  probabilistic attachment  of new vertices to leaves.
We  construct a  likelihood function of the leaves  based on the connectivity of the tree.
We take such  connectivity  to be  induced  by the  merging  of directed ordered paths from leaves to the root.
Combining the likelihood with an assigned prior  distribution leads to a posterior leaf distribution 
from which we sample attachment points for  new vertices.
We present computational examples of  such Bayesian tree  growth. 
Although the discussion is generic, the initial motivation for the paper is the concept of a distributed ledger, 
which may be regarded as a  time-ordered random tree that grows by probabilistic leaf attachment. 
\end{abstract}

\section{Introduction}
In the context of this paper, 
a {\it tree} is an object in graph theory. 
In particular, we study a {\it directed rooted tree} (vertices joined by directed edges where one vertex is the root)
that grows with time according to probabilistic rules. 
The motivation for such a study will be discussed 
below.
In the first instance, we give a simple  illustration of the growth of a directed rooted tree. 
 
 \begin{figure}[tbh]
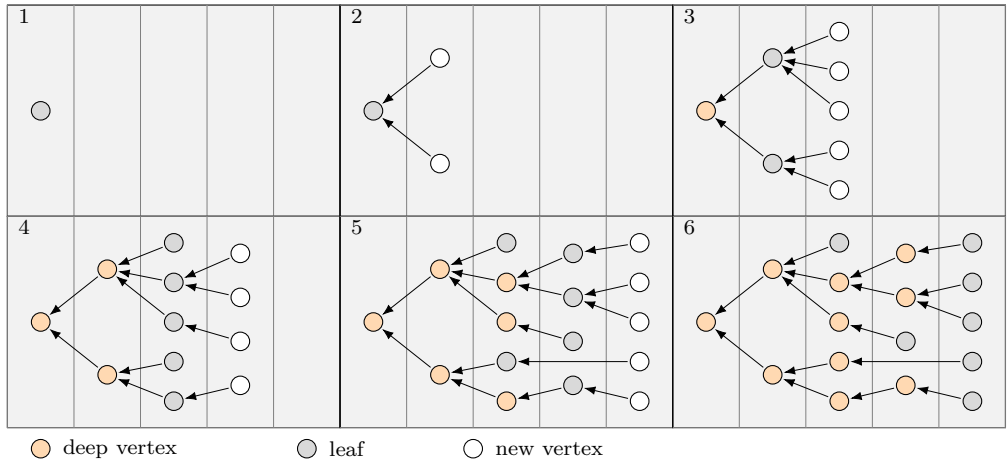

\centering
\include{Figure_tree_6frames}
\caption{
Tree grows sequentially from  1 to 6 through leaf attachment (appending a new vertex to the end of a branch).
}
\label{fig:treegraph}
\end{figure}

We adopt an approach to graph theory where visual representation often takes precedence over formal description.
Accordingly, Figure~\ref{fig:treegraph} shows a sequence of `snapshots' of a tree as it  grows
through attachment of new vertices to existing vertices
(time increases  to the right in each snapshot and
a directed edge from one vertex to another represents attachment of the former to the latter). 
The tree has two  vertex types: 
\\
a)  {\it deep} vertex  with one or more incoming attachments \\ 
b) {\it leaf}   
with no incoming attachments (root is also taken to be a leaf at the start).\\
A new vertex, created in the latest time interval, becomes  part of the tree only after attaching to an  existing vertex.
The growth from a given root obeys the following rules: 
\begin{labeling}{Rule 5:}
\setlength{\topsep}{-4pt}
\setlength{\itemsep}{0pt}
\setlength{\parskip}{0pt}
\setlength{\parsep}{0pt} 

\item[Rule~1:] Only new vertices may issue attachments, existing 
 vertices are `inactive'.
\item[Rule~2:] A new vertex  issues exactly one attachment (this can be generalised). 
\item[Rule~3:] {\it Leaf attachment}: a new vertex may only attach to a leaf.
 \item[Rule~4:] A leaf can receive multiple attachments from different  new vertices created in the same time interval.

 \item[Rule~5:] New vertices select attachment targets from a probability distribution over leaves.
\end{labeling}
It follows from the rules that attachments always point backward in time. 
The object of study is thus a time-ordered directed random tree that grows by probabilistic leaf attachment.

There are two logically distinct processes at play. 
The first is the  creation of a number of new vertices  in the current time interval. 
We can take this number  to be drawn from a  Poisson distribution, whose rate may be time-dependent.
However, this is not a fundamental design feature.
 Alternative vertex creation mechanisms, deterministic or probabilistic,   are  permissible.
 The second process is the attachment of  newly created vertices  to leaves of the  existing tree. 
 It is the latter process that is of interest here.
 Accordingly, the problem statement is the following:
 
\begin{tcolorbox}[ fonttitle=\bfseries, title=Problem Statement, colback=gray!10!white, colframe=gray!80!black]
This paper is devoted to {\sffamily Rule~5} --  
choosing a probability distribution over leaves from which new vertices select their target attachments.
Attaching entirely \textquote{at random} is equivalent to choosing  a uniform leaf distribution, 
regardless of the tree structure. 
The novelty of the paper is  a Bayesian scheme to update any prior choice of leaf distribution,  uniform or otherwise,
by taking the historical  connectivity of the tree into account.
\end{tcolorbox}

We first present a motivation for the study. 

\subsection{Motivation}
\label{sec:motivation}
The  initial inspiration for this paper is an innovation in computer science known as  a {\it distributed ledger}. 
It is used to enable secure transactions  over  peer-to-peer networks
(distributed  networks with  no  central  data repository or  coordinating authority). 
There is no  trusted party to 
keep a record of  transactions under secure `lock and key'. 
Yet there remains a need for a record-keeping mechanism that can be updated but not retroactively modified.
This record must necessarily also be distributed amongst network peers  in the absence of central storage or coordination.
A transaction originated by one peer  must be broadcast to all peers in order to maintain consistency of 
the distributed record keeping mechanism.
This  is what a distributed ledger seeks to achieve.

In the language of computer science, the ledger must be an immutable \textquote{append-only}  data structure.
Such a data structure might logically consist of  a time-ordered  sequence of transactions, updated by adding the latest transaction
 to the end of the sequence.
 Immutability requires a mechanism to ensure that  a negligent or  hostile actor  can neither  insert  a new transaction 
 before the end, nor  delete or modify an existing entry.

Nakamoto~\cite{Nakamoto}  introduced a distributed ledger to record the exchange of a digital currency called {\it bitcoin}
on a peer-to-peer  network -- the bitcoin network.
Peers gather  transactions  into blocks which are  chained  together with the aid of cryptography to attain immutability.
This immutable but verifiable cryptographic sequence of blocks has come  to be known as the  {\it blockchain}.

A  sequence  is not the only imaginable  append-only data structure. 
A tree  where  new entries are appended to leaves is an alternative generalisation. 
Indeed, blockchain is, in fact, derived from a tree. 
Peers can receive multiple blocks at the same (discrete) time that were  broadcast from different parts of the network. 
A strict  time-order acceptance rule no longer applies, 
implying simultaneous acceptance of such blocks so that the ledger branches into a tree.
When one sequence from root to leaf  becomes longer than the others, 
Nakamoto chooses it as the effective blockchain, to which new blocks must attach. 
 
 Sompolinsky and Zohar~\cite{SompZohar1} suggested an alternative to the  longest sequence that 
\textquote{selects at each fork in the chain the heaviest subtree rooted at the fork}. 
As discussed in~\cite{SompZohar1}, 
blocks associated with  greater branching enhance  security, amongst other advantages. 

The \textquote{longest chain} and \textquote{heaviest subtree}  deterministic rules select a single leaf to 
which new blocks must attach, abandoning all other leaves. 
A less extreme approach is to define a leaf probability distribution that assigns non-zero probability to every leaf.
Attachment points for new blocks can then be sampled probabilistically  from this leaf distribution 
rather than deterministically prescribed.
In this way, the whole tree continues to be the active distributed ledger rather than a selected chain leading to a single leaf.

Popov~\cite{Popov1} adopted a  variant on the method of Sompolinsky and Zohar  
involving  weighted random walks from root to leaf, with  probabilistic branching based on relative subtree weights. 
The terminal leaf of the walk then becomes the point of attachment for new blocks.
Hence the tree grows  probabilistically without abruptly abandoning all chains but one, as deterministic rules do.
(To be precise, Popov immediately generalised the tree to a graph by allowing a new vertex  to make multiple attachments,
in an approach aimed at the internet-of-things where  vertices represent individual transactions rather than blocks of transactions.)

Popov's weighted random walk  aims to generate a sample from a leaf distribution without explicitly constructing it.
In this paper, we construct the explicit probability distribution of leaves based on the connectivity of the tree and then 
sample attachment points  for new vertices from it.
More significantly, we present a novel approach that involves exploring suitably weighted directed paths from leaf to root 
rather than branching from root to leaf.

Having drawn inspiration from blockchain, we may set it aside to discuss random trees by
probabilistic leaf attachment in the abstract.
The  generic topic  is of mathematical and computational interest  in its own right, 
whether or not  it finds application in alternative distributed ledger design 
or other context not envisaged here.

We conclude by noting that we have consciously glossed over many challenges arising from the  
absence of a central authority in a peer-to-peer network.
For example, there are variable broadcast delays so that, at any instant, peers may hold different snapshots
of the ledger.
This requires consensus rules on such issues as what constitutes the longest chain or heaviest subtree sequence at any stage.
Furthermore, aside from attempted cryptanalysis, distributed ledgers have to contend with  a slew of  hostile attacks.
 For instance, Nakamoto 
 used the Gambler's Ruin problem  (Feller~\cite{Feller1}, p342) to model a particular attack scenario.
A comprehensive discussion of distributed ledgers would involve computer science,  discrete mathematics 
(graph theory, cryptography), probability theory and game theory.

\subsection{Context and Scope}
\label{sec:context}

We have discussed the distributed ledger as an example of  a random tree that grows by probabilistic leaf attachment. 
The tree  can be generalised to a random graph if new vertices are allowed to make multiple  attachments.
We consciously  limit the scope here to a tree, from which much can be learned before tackling the more general graph.

There is another context  involving random graphs that is worth commenting on to avoid 
potential misinterpretation of the theme of this paper.
The bitcoin network 
referred to earlier is an example of a real-world network,
much like biophysical networks, the internet,  social networks, scholarly citation networks {\it etc.}
In a  field  of study  known  as  network science,
random graphs are used as mathematical models of complex real-world networks 
(Barab\'{a}si~\cite{Barabasi}, van der Hofstad~\cite{vdHofstad}, Newman~\cite{Newman}).
Such networks are typically dynamic, evolving through structural reconfiguration or growth.
The  random graph models must  correspondingly reflect such evolution in some  probabilistic sense.

Pioneering work on random graphs considered a fixed number of vertices with a probability distribution over 
(i) the number of edges  for classes of graphs (Erd\H{o}s and R\'{e}nyi~\cite{ErdosRenyi}),
or (ii) placement of edges for a given graph (Gilbert~\cite{Gilbert}).
However, real networks often grow in time through attachment of new vertices.
For example, a  new member of a social  network  may prefer to befriend members who are already well-connected to other members.
To capture such preference in the  random graph model,  
a  new vertex may attach to a current vertex with probability proportional to the {\it degree} of the latter
(the degree of a vertex is the number of  edges connecting it to other vertices:  a  measure of the vertex's connectivity).
Barab\'{a}si and Albert~\cite{BarabasiAlbert} studied such  degree-biased attachment for undirected graphs, 
which  has come to be known as  \emph{preferential attachment}.
The concept was first described by Yule~\cite{Yule} and Simon~\cite{Simon} and it is associated with the Yule-Simon distribution.
Bollob\'{a}s {\it et al.}~\cite{Bollobas} extended preferential attachment to directed graphs,
where edges have a one-directional meaning, so that 
a vertex has two degree types:  in-degree (number of incoming edge) and out-degree (number of outgoing edges).

 Generalisations include attachment probability proportional to a
 nonlinear function of degree (Krapivsky {\it et al.}~\cite{Krapivsky}) 
and degree weighted  by, say, assigning an additional  `fitness'  variable to vertices (Bianconi and Barab\'{a}si~\cite{Bianconi}).
Preferential attachment 
has received much interest because it leads to  \textquote{scale-free} graphs,
{\it i.e.}\ power-law degree distributions at  very large scale. 
The  prevalence of scale-free behaviour in  complex real-world networks compared to their random graph counterparts 
is a subject of  debate in network science (Broido and Clauset~\cite{BroidoClauset}).

The bitcoin network is an instance of a real-world network that  can be modelled as a  random graph. 
Kondor {\it et al.}~\cite{Kondor} explored the `rich-get-richer' consequence of preferential attachment for the bitcoin network.
Javarone and Wright~\cite{Javarone}  suggested the Bianconi-Barab\'{a}si random graph representation.
The bitcoin network model is logically distinct from the distributed  ledger that keeps a record of transactions issued by the network.

A network science model   that resembles  the problem of this paper is the  directed  time-ordered  graphical model 
of citations of scholarly publications.
It was  initially studied by  Price~\cite{Price},  who referred to preferential attachment as {\it cumulative advantage}.
A new vertex  represents a new publication and an attachment to an existing  vertex  represents citation of an older publication.
Conceptually, this might  be regarded as a more general variant of our problem,
where a new vertex may make multiple attachments to existing  vertices at any depth  of the graph.
However, beside restriction to leaf attachment,  our context differs  fundamentally from that  of citation modelling 
or  any other network science study. 

Given a random graph model of a real-world network,  the natural question of network science is:
\textquote{Do real-world complex networks actually behave like that?}.
The question of interest here is:
\textquote{Does  the random tree/graph behave according to design specifications?}.
This is  more of an engineering  problem  (designing and testing a product with prescribed characteristics)  
 than a scientific problem (modelling an observable  phenomenon).
It is akin to designing an aircraft and  testing a prototype's  flight capabilities against expectation.
This may well  be inspired by observing birds in flight but  it need not try to mimic them through a mathematical model of their prowess, 
including  properties like flock intelligence that might emerge at `large scale'.
Quite separately, a complex network of flight paths and destinations woven by commercial use of  aircraft might be modelled by 
random graphs,  along with questions about the  large scale behaviour of model {\it versus} reality.

The upshot of all this is that we shall not feel obliged to explore the asymptotic behaviour of  our leaf attachment scheme.

\subsection{Trees Everywhere}
\label{sec:trees}

 For completeness, we note that a  directed rooted tree is a common representation of a hierarchy of concepts and actions 
in  practically any field.
The tree typically grows by  branching,  sprouting new vertices like a natural tree, 
rather than by attachment of separately created vertices to the existing tree.
The branching process can be probabilistic or deterministic.

An early example of probabilistic branching is the Galton-Watson tree in genealogy.
First studied  in the 1800s (Bienaym\'{e}~\cite{Bienayme}, Galton and Watson~\cite{GaltonWatson}), 
it is  well-documented in subsequent literature on probability and trees
({\it e.g.}\ Drmota~\cite{Drmota}, Lyons and Peres \cite{LyonsPeres}).

A simple example of  deterministic branching  is the binary search tree of computer science.
It is, in turn, an  instance of a decision tree, which finds much use in 
machine learning ({\it e.g.}\ Quinlan~\cite{Quinlan} and a wealth of more recent literature).

Unlike natural trees, abstract trees can also  grow backward through merging of  paths from leaf to root.
An example is the Merkle tree (Merkle~\cite{Merkle}), which starts with  leaves,
 each  being a cryptographic hash function of specified data.
Proceeding from leaf to root, each inner  vertex is, in turn,  a  hash of two preceding vertices (for a binary implementation),
 terminating in a hash of the whole tree at the root.
Since cryptographic hash functions are one-way (non-invertible), the  Merkle tree can only grow from leaf to root.
It plays a fundamental role in blockchain design for cryptographic compression of  transactions in a  block 
(not to be confused with the tree structure that arises as the blockchain grows).

While trees can be studied in the abstract in graph theory, they also find application as a representational tool in mathematics.
For example, the Collatz tree  is a representation of  the Collatz conjecture of number theory.
A Quanta article~\cite{Quanta} on a recent advance  toward a proof of the conjecture by  
Tao~\cite{Tao} opened with an animation of a growing  Collatz tree.
The  animation, due to   Davies~\cite{Davies}, displays the tree in outward-growing concentric  layers.
The Collatz conjecture  is normally formulated  so that the Collatz tree (like the Merkle tree)  
 grows inward, governed by a simple integer function, from any starting vertex (integer $n>1$) to the root ($n=1$).
 But (unlike the Merkle tree), this function is invertible so that the Collatz tree can also
grow by  branching  from the root.


\subsection{Novelty}
\label{sec:novel}

There are two steps to the growth of our tree: \\
1) Constructing the leaf distribution given the current snapshot of the tree \\
2) Growing the tree by attachment of new vertices to leaves based on the leaf distribution. 

It may seem natural to expect  the leaf distribution to be  constructed by branching from the root.
 On the other hand, the Merkle and Collatz trees -- deterministic though they may be -- 
  inspire the thought of inward exploration of the snapshot, 
 gathering  knowledge about the tree from leaf to root instead.

The key insight here is that  we can indeed   traverse the tree  in the inward direction along directed paths from leaf to root.
In so doing, we learn about the history of the tree in an appropriate probabilistic sense. 
Then we can use Bayes' rule to infer the leaf distribution, without further need to branch outward from root to leaf.

To our awareness, such a conceptual  framework   is  novel.
Bayesian analysis has   been applied to decision trees 
for classification in machine learning (dating back to  Buntine~\cite{Buntine}),
but this is a different application  from that considered here.
Decision trees grow by branching rather than attachment, 
hence the question of constructing a leaf distribution to guide attachment does not arise.

\subsection{Structure of Paper}
The detail of the proposed path-based Bayesian method will be discussed in section~\ref{sec:bayes}.
For completeness, we will also explore a degree-based branching  approach in section~\ref{sec:branch}.
We present computational examples in section~\ref{sec:comp}.

\section{Probabilistic Tree Growth}
\label{sec:bayes}

Let $\mathscr L$ be the set of leaves and let  $\cH$ denote the history of the  tree,  {\it i.e.}\  the observable data about the tree.
We proceed as follows:
\begin{labeling}{Likelihood:}
\setlength{\topsep}{-4pt}
\setlength{\itemsep}{0pt}
\setlength{\parskip}{0pt}
\setlength{\parsep}{0pt} 
\item[Prior:] Assign an initial leaf distribution $\Pr(\ell)$ for the discrete variable $\ell\in \mathscr L$ 
without  prior knowledge of the  history of  the  snapshot of the tree.
In the absence of any other information, the  natural choice  is a uniform prior.
If we choose to ignore the tree's history, attachment points for new vertices may, without further ado,  
be sampled from this prior.
Else proceed to the next step. 
\item[Likelihood:] Learn about  the history  of the tree by  exploring it backward  from leaf to  root,
{\it i.e.}\ construct the likelihood $\Pr(\cH|\ell)$.
\item[Posterior:] Construct the  posterior leaf distribution $\Pr(\ell|\cH)$ by making use of the product and sum rules of probability theory
\begin{alignat}{3}
& \Pr(\ell, \cH) &&= \Pr(\cH | \ell)\Pr(\ell) = \Pr(\ell | \cH) \Pr(\cH)  
\label{eq:joint} \\
\implies\;&
\Pr(\ell | \cH) &&= \Pr(\cH|\ell)\Pr(\ell) / \Pr(\cH) \qquad \textrm{Bayes' rule}
\label{eq:bayes} \\
\textrm{where}\; &
 \Pr(\cH) &&= \sum_{\ell\in\mathscr L}  \Pr(\ell, \cH)  = \sum_{\ell\in\mathscr L} \Pr(\cH|\ell)\Pr(\ell)  
 \label{eq:PrH} 
 \end{alignat}

\item[Growth:] Given $n$ new vertices where $n$ is generated from a Poisson distribution (say), 
sample the $n$  attachment points  for these new vertices from $\Pr(\ell|\cH)$.
\end{labeling}

\begin{remark} 
Despite the presentation order, constructing the likelihood is  independent of the choice of prior.
Both are  needed to construct the posterior.
\end{remark} 
\begin{remark} 
 In general, the use of the posterior depends on the application context. 
The  sole aim  here is to sample attachment  targets from the posterior in order to grow the tree.
This contrasts with the  common objective of Bayesian analysis where the posterior is used for parameter estimation or classification.
\end{remark} 
\begin{remark} 
$\mathscr L$ is a  set of any  distinct leaf labels, such as  simple leaf number.
We strike no distinction here between events of a sample space and corresponding random variables.
\end{remark} 
\begin{remark} 
 There is  a single history of the tree. 
Hence $\Pr(\cH|\ell)$ must be understood not as  a probability distribution over multiple instances of history, 
but  as a function of the  leaf  variable  $\ell$, known as the {\rm \sffamily likelihood} function. 
Similarly, $\Pr(\cH)$ is a constant rather than a distribution over different instances of $\cH$.
The only probability distributions involved are  the prior $\Pr(\ell)$ and the posterior  $\Pr(\ell|\cH)$.
\end{remark} 

Construction of the likelihood depends on  design objectives.
Our preference in this paper 
 is for  new vertices to attach to leaves that are directly or indirectly connected 
 to multiple branches of the current tree. 
Equivalently, we aim to discourage the growth of isolated chains of vertices.
This, in turn, guides our construction of $\Pr(\cH|\ell)$.
It is thus appropriate at this point to be explicit about the meaning of $\cH$.  
We shall make repeated use of the following  concept:
\begin{definition}
There is unique  sequence of attachments from any  leaf $\ell\in\mathscr L$ to the root  $r$ 
that we refer to as a {\rm \sffamily directed ordered path}, 
denoted by $\ell\twoheadrightarrow r$.
 The sequence of connected vertices along the path is strictly ordered from later to earlier in time.
\end{definition}

We use $y\rightarrow x$ to denote a direct attachment of vertex $y$ to vertex $x$.
Any direct attachment necessarily  lies on at least one path and possibly multiple merged paths. 
The more  merging there is the better connected the tree is.
How we use this  to construct a likelihood 
depends on how we interpret the paths as quantitative data.
To that end, we endow an attachment and a path with a weight as follows:
\begin{definition}
The {\rm \sffamily weight} (or {\rm \sffamily multiplicity}) of an attachment  $y\rightarrow x$, denoted by $|y\rightarrow x|$,  
is the number of  paths from the set $\mathscr L$ of leaves  to the root $r$ that  pass through $y\rightarrow x$.
\end{definition}
\begin{definition}
The {\rm \sffamily weighted length} (or just  {\rm \sffamily weight}) of a path $\ell\twoheadrightarrow r$  is the sum of weights of all 
intermediate attachments between $\ell$ and $r$. We denote this weight by $|\ell\twoheadrightarrow r|$. 
\end{definition}
\begin{figure}[tbh]
\centering
\begin{tikzpicture}[xscale=0.6, yscale=0.6,
Myfill/.style ={fill=#1!30, draw=black},
Myfill/.default=orange]

\newcommand{\drawEdge}[2]
{
\draw[Myarrow] let \p1 =  ($(#2) - (#1)$), \n1 = {atan2(\y1,\x1)}, \p2 = ($(\n1:\offset)$) in  ($(#1)+(\p2)$) -- ($(#2)-(\p2)$)
}
\newcommand*{\nx}{8}%
\newcommand*{\ny}{8}%
\newcommand*{\xo}{0}%
\newcommand*{\yo}{\ny/2}%
\newcommand*{\h}{0.1}%
\newcommand*{\xstep}{2.75}%
\newcommand*{\ystep}{1.0}%
\newcommand*{\radius}{0.55}%
\newcommand*{\radiusa}{0.45}%
\newcommand*{\offset}{\radius+\h}%
\newcommand*{\size}{\small}%
\newcommand*{\sizea}{\footnotesize}%
\newcommand*{\sizeb}{\scriptsize}%
\newcommand*{\tipcol}{gray}%
\newcommand*{\newcol}{white}%
\newcommand*{\backhue}{10}%


\newcommand*{\num}{4}
\newcommand*{\hya}{\ny/5}%
\newcommand*{\hyb}{\ny/5}%
\newcommand*{\hyc}{\ny/5}%

\coordinate (root) at (\xo,\yo);
\coordinate (x1) at ($(root)+(\xstep,-1.0*\hya*\ystep)$);
\coordinate (x2) at ($(x1)+(0,\hya*\ystep)$);
\coordinate (x3) at ($(x2)+(0,\hya*\ystep)$);

\coordinate (y1) at ($(root)+(2*\xstep,-1.5*\hyb*\ystep)$);
\coordinate (y2) at ($(y1)+(0,\hyb*\ystep)$);
\coordinate (y3) at ($(y2)+(0,\hyb*\ystep)$);
\coordinate (y4) at ($(y3)+(0,\hyb*\ystep)$);

\coordinate (z1) at ($(root)+(3*\xstep,-1.25*\hyc*\ystep)$);
\coordinate (z2) at ($(z1)+(0,\hyc*\ystep)$);
\coordinate (z3) at ($(z2)+(0,\hyc*\ystep)$);
\coordinate (z4) at ($(z3)+(0,\hyc*\ystep)$);

\draw [fill=gray!\backhue]  ($(root)-(0.5*\xstep,\yo)$) rectangle +($(\num*\xstep, \ny)$);
\draw [draw=black]  ($(root)-(0.5*\xstep,\yo)$) rectangle +($(\num*\xstep, \ny)$);

\filldraw [Myfill] (root) circle (\radius) node[font=\size]{$r$};
\filldraw [Myfill] (x1) circle (\radius) node[font=\size]{$d_1$};
\filldraw [Myfill] (x2) circle (\radius) node[font=\size]{$d_2$};
\filldraw [Myfill] (x3) circle (\radius) node[font=\size]{$d_3$};

\filldraw [Myfill=\tipcol] (y1) circle (\radius) node[font=\size]{$\ell_1$}; 
\filldraw [Myfill] (y2) circle (\radius) node[font=\size]{$d_4$}; 
\filldraw [Myfill] (y3) circle (\radius) node[font=\size]{$d_5$}; 
\filldraw [Myfill] (y4) circle (\radius) node[font=\size]{$d_6$}; 

\filldraw [Myfill=\tipcol] (z1) circle (\radius) node[font=\size]{$\ell_2$}; 
\filldraw [Myfill=\tipcol] (z2) circle (\radius) node[font=\size]{$\ell_3$}; 
\filldraw [Myfill=\tipcol] (z3) circle (\radius) node[font=\size]{$\ell_4$}; 
\filldraw [Myfill=\tipcol] (z4) circle (\radius) node[font=\size]{$\ell_5$}; 

\drawEdge{x1}{root} node[font=\sizeb, midway, fill=gray!\backhue]{$1$};
\drawEdge{x2}{root} node[font=\sizeb, midway, fill=gray!\backhue]{$3$};
\drawEdge{x3}{root} node[font=\sizeb, midway, fill=gray!\backhue]{$1$};
\drawEdge{y1}{x1} node[font=\sizeb, midway, fill=gray!\backhue]{$1$};
\drawEdge{y2}{x2} node[font=\sizeb, midway, fill=gray!\backhue]{$2$};
\drawEdge{y3}{x2} node[font=\sizeb, midway, fill=gray!\backhue]{$1$};
\drawEdge{y4}{x3} node[font=\sizeb, midway, fill=gray!\backhue]{$1$};
\drawEdge{z1}{y2} node[font=\sizeb, midway, fill=gray!\backhue]{$1$};
\drawEdge{z2}{y2} node[font=\sizeb, midway, fill=gray!\backhue]{$1$};
\drawEdge{z3}{y3} node[font=\sizeb, midway, fill=gray!\backhue]{$1$};
\drawEdge{z4}{y4} node[font=\sizeb, midway, fill=gray!\backhue]{$1$};

\end{tikzpicture}
\caption{Tree with root $r$, deep vertices $\{d_i\}$ and leaves $\{\ell_i\}$.
The attachment label is the number of  paths from leaf to root  passing through that attachment}
\label{fig:paths1}
\end{figure}
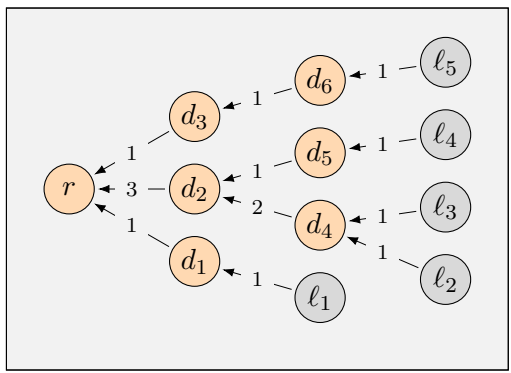

\begin{figure}[tbh]
\centering
\begin{tikzpicture}[xscale=0.6, yscale=0.6,
Myfill/.style ={fill=#1!30, draw=black},
Myfill/.default=orange]

\newcommand{\drawEdge}[2]
{
\draw[Myarrow] let \p1 =  ($(#2) - (#1)$), \n1 = {atan2(\y1,\x1)}, \p2 = ($(\n1:\offset)$) in  ($(#1)+(\p2)$) -- ($(#2)-(\p2)$)
}
\newcommand*{\nx}{8}%
\newcommand*{\ny}{8}%
\newcommand*{\xo}{0}%
\newcommand*{\yo}{\ny/2}%
\newcommand*{\h}{0.1}%
\newcommand*{\xstep}{2.75}%
\newcommand*{\ystep}{1.0}%
\newcommand*{\radius}{0.55}%
\newcommand*{\radiusa}{0.45}%
\newcommand*{\offset}{\radius+\h}%
\newcommand*{\size}{\small}%
\newcommand*{\sizea}{\footnotesize}%
\newcommand*{\sizeb}{\scriptsize}%
\newcommand*{\tipcol}{gray}%
\newcommand*{\newcol}{white}%
\newcommand*{\backhue}{10}%


\newcommand*{\num}{4}
\newcommand*{\hya}{\ny/5}%
\newcommand*{\hyb}{\ny/5}%
\newcommand*{\hyc}{\ny/5}%

\coordinate (root) at (\xo,\yo);
\coordinate (x1) at ($(root)+(\xstep,-1.0*\hya*\ystep)$);
\coordinate (x2) at ($(x1)+(0,\hya*\ystep)$);
\coordinate (x3) at ($(x2)+(0,\hya*\ystep)$);

\coordinate (y1) at ($(root)+(2*\xstep,-1.5*\hyb*\ystep)$);
\coordinate (y2) at ($(y1)+(0,\hyb*\ystep)$);
\coordinate (y3) at ($(y2)+(0,\hyb*\ystep)$);
\coordinate (y4) at ($(y3)+(0,\hyb*\ystep)$);

\coordinate (z1) at ($(root)+(3*\xstep,-1.25*\hyc*\ystep)$);
\coordinate (z2) at ($(z1)+(0,\hyc*\ystep)$);
\coordinate (z3) at ($(z2)+(0,\hyc*\ystep)$);
\coordinate (z4) at ($(z3)+(0,\hyc*\ystep)$);

\draw [fill=gray!\backhue]  ($(root)-(0.5*\xstep,\yo)$) rectangle +($(\num*\xstep, \ny)$);
\draw [draw=black]  ($(root)-(0.5*\xstep,\yo)$) rectangle +($(\num*\xstep, \ny)$);
\draw ($(root)+(-\xstep/2.5,0.9*\yo)$) node[font=\scriptsize]{1};

\filldraw [Myfill] (root) circle (\radius) node[font=\size]{$r$};
\filldraw [Myfill] (x1) circle (\radius) node[font=\size]{$d_1$};
\filldraw [Myfill] (x2) circle (\radius) node[font=\size]{$d_2$};
\filldraw [Myfill] (x3) circle (\radius) node[font=\size]{$d_3$};

\filldraw [Myfill=\tipcol] (y1) circle (\radius) node[font=\sizea]{$1|2$}; 
\filldraw [Myfill] (y2) circle (\radius) node[font=\size]{$d_4$}; 
\filldraw [Myfill] (y3) circle (\radius) node[font=\size]{$d_5$}; 
\filldraw [Myfill] (y4) circle (\radius) node[font=\size]{$d_6$}; 

\filldraw [Myfill=\tipcol] (z1) circle (\radius) node[font=\sizea]{$1|6$}; 
\filldraw [Myfill=\tipcol] (z2) circle (\radius) node[font=\sizea]{$1|6$}; 
\filldraw [Myfill=\tipcol] (z3) circle (\radius) node[font=\sizea]{$1|5$}; 
\filldraw [Myfill=\tipcol] (z4) circle (\radius) node[font=\sizea]{$1|3$}; 

\drawEdge{x1}{root} node[font=\sizeb, midway, fill=gray!\backhue]{$1$};
\drawEdge{x2}{root} node[font=\sizeb, midway, fill=gray!\backhue]{$3$};
\drawEdge{x3}{root} node[font=\sizeb, midway, fill=gray!\backhue]{$1$};
\drawEdge{y1}{x1} node[font=\sizeb, midway, fill=gray!\backhue]{$1$};
\drawEdge{y2}{x2} node[font=\sizeb, midway, fill=gray!\backhue]{$2$};
\drawEdge{y3}{x2} node[font=\sizeb, midway, fill=gray!\backhue]{$1$};
\drawEdge{y4}{x3} node[font=\sizeb, midway, fill=gray!\backhue]{$1$};
\drawEdge{z1}{y2} node[font=\sizeb, midway, fill=gray!\backhue]{$1$};
\drawEdge{z2}{y2} node[font=\sizeb, midway, fill=gray!\backhue]{$1$};
\drawEdge{z3}{y3} node[font=\sizeb, midway, fill=gray!\backhue]{$1$};
\drawEdge{z4}{y4} node[font=\sizeb, midway, fill=gray!\backhue]{$1$};

\coordinate (root) at (\num*\xstep,\yo);
\coordinate (x1) at ($(root)+(\xstep,-1.0*\hya*\ystep)$);
\coordinate (x2) at ($(x1)+(0,\hya*\ystep)$);
\coordinate (x3) at ($(x2)+(0,\hya*\ystep)$);

\coordinate (y1) at ($(root)+(2*\xstep,-1.5*\hyb*\ystep)$);
\coordinate (y2) at ($(y1)+(0,\hyb*\ystep)$);
\coordinate (y3) at ($(y2)+(0,\hyb*\ystep)$);
\coordinate (y4) at ($(y3)+(0,\hyb*\ystep)$);

\coordinate (z1) at ($(root)+(3*\xstep,-1.25*\hyc*\ystep)$);
\coordinate (z2) at ($(z1)+(0,\hyc*\ystep)$);
\coordinate (z3) at ($(z2)+(0,\hyc*\ystep)$);
\coordinate (z4) at ($(z3)+(0,\hyc*\ystep)$);

\draw [fill=gray!\backhue]  ($(root)-(0.5*\xstep,\yo)$) rectangle +($(\num*\xstep, \ny)$);
\draw [draw=black]  ($(root)-(0.5*\xstep,\yo)$) rectangle +($(\num*\xstep, \ny)$);
\draw ($(root)+(-\xstep/2.5,0.9*\yo)$) node[font=\scriptsize]{2};

\filldraw [Myfill] (root) circle (\radius) node[font=\size]{$r$};
\filldraw [Myfill] (x1) circle (\radius) node[font=\size]{$d_1$};
\filldraw [Myfill] (x2) circle (\radius) node[font=\size]{$d_2$};
\filldraw [Myfill] (x3) circle (\radius) node[font=\size]{$d_3$};

\filldraw [Myfill=\tipcol] (y1) circle (\radius) node[font=\sizea]{$1|1$}; 
\filldraw [Myfill] (y2) circle (\radius) node[font=\size]{$d_4$}; 
\filldraw [Myfill] (y3) circle (\radius) node[font=\size]{$d_5$}; 
\filldraw [Myfill] (y4) circle (\radius) node[font=\size]{$d_6$}; 

\filldraw [Myfill=\tipcol] (z1) circle (\radius) node[font=\sizea]{$1|6$}; 
\filldraw [Myfill=\tipcol] (z2) circle (\radius) node[font=\sizea]{$1|6$}; 
\filldraw [Myfill=\tipcol] (z3) circle (\radius) node[font=\sizea]{$1|3$}; 
\filldraw [Myfill=\tipcol] (z4) circle (\radius) node[font=\sizea]{$1|1$}; 

\drawEdge{x1}{root} node[font=\sizeb, midway, fill=gray!\backhue]{$1$};
\drawEdge{x2}{root} node[font=\sizeb, midway, fill=gray!\backhue]{$3$};
\drawEdge{x3}{root} node[font=\sizeb, midway, fill=gray!\backhue]{$1$};
\drawEdge{y1}{x1} node[font=\sizeb, midway, fill=gray!\backhue]{$1$};
\drawEdge{y2}{x2} node[font=\sizeb, midway, fill=gray!\backhue]{$2$};
\drawEdge{y3}{x2} node[font=\sizeb, midway, fill=gray!\backhue]{$1$};
\drawEdge{y4}{x3} node[font=\sizeb, midway, fill=gray!\backhue]{$1$};
\drawEdge{z1}{y2} node[font=\sizeb, midway, fill=gray!\backhue]{$1$};
\drawEdge{z2}{y2} node[font=\sizeb, midway, fill=gray!\backhue]{$1$};
\drawEdge{z3}{y3} node[font=\sizeb, midway, fill=gray!\backhue]{$1$};
\drawEdge{z4}{y4} node[font=\sizeb, midway, fill=gray!\backhue]{$1$};

\end{tikzpicture}
\caption{As in Figure~\ref{fig:paths1}, with unnormalised prior$|$posterior values on leaves for two likelihood choices:
(1) Global -- based on weighted path length, (2) Local --  based on individual attachment weights.}
\label{fig:paths2}
\end{figure}
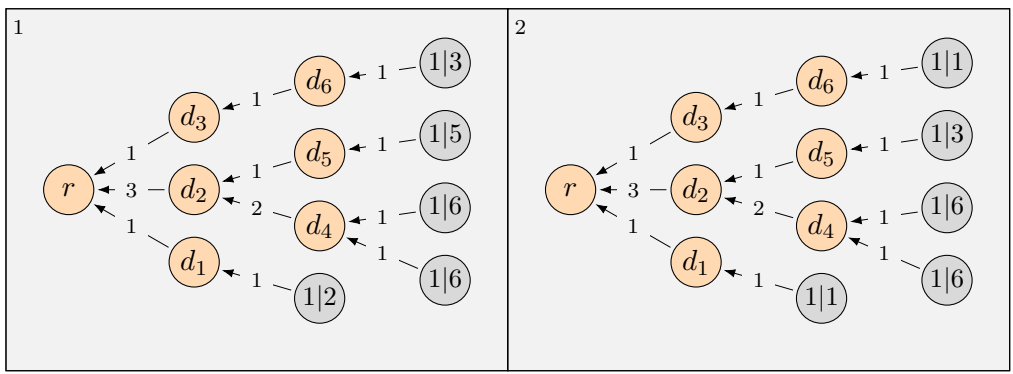

Figure~\ref{fig:paths1} shows root $r$, deep vertices $\{d_1,\ldots,d_6\}$ and leaves $\mathscr L=\{\ell_1,\ldots,\ell_5\}$.
Each  leaf initiates a path to $r$, {\it e.g.}\ path $\ell_2\twoheadrightarrow r \equiv \ell_2\rightarrow d_4\rightarrow d_2\rightarrow\ r $.
The label on each attachment is its weight as defined above.

\subsection{Global Interpretation}
\label{sec:global}
We identify $\Pr(\cH|\ell)$ with a global attribute of the path $\ell\twoheadrightarrow r$, {\it viz.}\ the path length: 
\begin{conditional}
The conditional probability of the path $\ell\twoheadrightarrow r$ given $\ell$ is taken to be the weighted path length
\begin{equation}
\Pr(\cH|\ell) = |\ell\twoheadrightarrow r|
\label{eq:condprob1}
\end{equation}
\end{conditional}

Hence the likelihood for the example of Figure~\ref{fig:paths1} is
\begin{equation}
\begin{alignedat}{3}
\Pr(\cH|\ell_1) &= 1+1 &&= 2 \\
\Pr(\cH|\ell_2) 
&= 3+2+1 &&= 6  \\
\Pr(\cH|\ell_3) 
&= 3+2+1 &&= 6  \\
\Pr(\cH|\ell_4) 
&= 3+1+1 &&= 5  \\
\Pr(\cH|\ell_5) 
&= 1+1+1 &&= 3 
\end{alignedat}
\label{eq:lhood1}
\end{equation}
We have made no assumptions about the prior $\Pr(\ell)$ 
and we can accommodate any choice.
For a uniform prior,  
 the posterior is  $\Pr(\ell|\cH)\propto\Pr(\cH|\ell)$.
The leaves in frame~1 of Figure~\ref{fig:paths2} show the prior on the left and the posterior on the right, 
omitting normalisation in each case.
The shorter isolated chain (terminating on leaf $\ell_1$) has lowest probability, followed by the longer one ($\ell_5$). 
Highest probability goes to $\ell_2$ and $\ell_3$, which exhibit the most merging of paths between leaf and root.
This is in keeping with the objective expressed earlier that the posterior must favour attachment to well-connected leaves 
and discourage isolated chains.


\subsection{Local Interpretation}
\label{sec:local}
We take an individual attachment as the  basic data object,
so that we describe a path $\ell\twoheadrightarrow r$ explicitly in terms of its  attachments. 
In Figure~\ref{fig:paths1}, $\ell_2\twoheadrightarrow r \equiv \ell_2\rightarrow d_4\rightarrow d_2\rightarrow r$.
In this instance, we write  $\Pr(\cH|\ell_2)$  as $\Pr(r,d_2,d_4|\ell_2)$ which, by the product rule, may be written as
\begin{equation}
\Pr(\cH|\ell_2) \equiv  \Pr(r,d_2,d_4|\ell_2) = \Pr(r|d_2)\Pr(d_2|d_4)\Pr(d_4|\ell_2) 
 \label{eq:markov}
\end{equation}
with analogous expressions for the other 4 paths from $\mathscr L$ to $r$.
The decomposition~(\ref{eq:markov}) follows from the Markov property that a vertex on a specified path is conditionally independent 
of vertices that {\it indirectly} attach to it,  given the vertex that {\it directly} attaches to it on that path.
Succinctly,  $\Pr(r|d_2,d_4,\ell_2) = \Pr(r|d_2)$, for example.

\begin{conditional}
The conditional probability of  $x$ given  $y$, where $y$ directly attaches to $x$ 
is taken to be the  attachment weight
\begin{equation}
\Pr(x|y)=|y\rightarrow x|
\label{eq:condprob}
\end{equation}
\end{conditional}
Hence the likelihood for the example of Figure~\ref{fig:paths1} is
\begin{equation}
\begin{alignedat}{4}
\Pr(\cH|\ell_1) &\equiv  \Pr(r,d_1|\ell_1)         &&= \Pr(r|d_1)\Pr(d_1|\ell_1)                    &&= 1\times 1              &&= 1 \\
\Pr(\cH|\ell_2) &\equiv  \Pr(r,d_2,d_4|\ell_2)  &&= \Pr(r|d_2)\Pr(d_2|d_4)\Pr(d_4|\ell_2) &&= 3\times 2\times 1 &&= 6 \\
\Pr(\cH|\ell_3) &\equiv  \Pr(r,d_2,d_4|\ell_3)  &&= \Pr(r|d_2)\Pr(d_2|d_4)\Pr(d_4|\ell_3) &&= 3\times 2\times 1 &&=  6 \\
\Pr(\cH|\ell_4) &\equiv  \Pr(r,d_2,d_5|\ell_4)  &&= \Pr(r|d_2)\Pr(d_2|d_5)\Pr(d_5|\ell_4) &&= 3\times 1\times 1 &&=  3 \\
\Pr(\cH|\ell_5) &\equiv  \Pr(r,d_3,d_6|\ell_5)  &&= \Pr(r|d_3)\Pr(d_3|d_6)\Pr(d_6|\ell_5) &&= 1\times 1\times 1 &&= 1 
\end{alignedat}
\label{eq:lhood}
\end{equation}

Frame~2 of Figure~\ref{fig:paths2} shows the prior and posterior.
The shape of the posterior is  indeed as intended, with isolated chains having the same lowest probability regardless of length.

\subsection{Discussion}
In summary, the history $\cH$ of the tree is the set of paths  $\{\ell\twoheadrightarrow r: \ell\in \mathscr L\}$ 
 from leaves to root, where a path consists of connected attachments from $\ell$ to $r$.
An attachment is weighted by the number of coincident paths  sharing that attachment.
The likelihood $\Pr(\cH|\ell)$  depends  only on the path $\ell\twoheadrightarrow r$ 
or,  more precisely, on some quantitative attribute thereof.
In the global interpretation, we take this attribute  to be the sum of weights of direct attachments 
along the path ({\it i.e.}\ the weight of the path), which we directly equate to $\Pr(\cH|\ell)$.
In the local interpretation, we take the weight of an  attachment to be the conditional probability between the two vertices 
associated with the attachment.  
Then $\Pr(\cH|\ell)$ is the product of such pairwise conditionals along the path.

The path-based  characterisation of a tree  generalises  to a time-ordered directed  graph,
where a vertex may make multiple attachments.
Hence, there may be multiple directed paths from any leaf to the root, 
where the paths may  merge with other paths as well as split into several  subsidiary paths.
In our view, characterising a directed  ordered graph in terms of directed ordered paths is more natural and useful 
than describing it as a directed acyclic graph (DAG), as per standard practice.
The acyclic property is a consequence of being directed and  ordered rather than a fundamental  attribute.
It is  like describing time evolution by saying that history does not repeat itself.
True though this may be (except in a purely figurative sense), it is not self-evidently useful.

 Karrer and Newman~\cite{KarrerNewman}  similarly commented that the term DAG  is
\textquote{perhaps slightly misleading, focusing our attention, as it does, 
on the acyclic property rather than the more fundamental ordering}.
While conceding that \textquote{directed ordered graph}  would be unlikely to find wide adoption,
they explicitly \textquote{incorporate an underlying ordering of the vertices that then drives the acyclic structure}.
Thus they described their model as  a \textquote{random ordered graph}.

In their discussion of the citation graph, Evans {\it et al.}~\cite{Evans} refer to a  fundamental \textquote{arrow of time} 
because a document can only cite documents older than itself.
This necessarily means that there can be no directed cycles (closed loops) in the graph as this can only arise from an earlier document paradoxically citing a later one.
Hence the graphical model of a citation network  is an instance of a DAG.
Although the concept of ordering is more generally applicable than time-ordering, 
we may implicitly take DAG -- whenever we use the term -- to be  a shorthand for \textquote{directed arrow-of-time graph}.

Our contention is that, at least for our leaf attachment problem, directed ordered paths  offer the natural way 
to characterise directed ordered graphs, or to make explicit use of the arrow-of-time.
The paths  may then  be endowed with probabilities, 
where the basic probabilistic construct is the conditional probability between two vertices
 rather than the probability of a vertex.
 In fact, the leaves are the only vertices that make up the `parameter space' and are thus endowed with prior and posterior probabilities.
 All other  vertices are part of the likelihood.
 
That said, as noted in section~\ref{sec:context}, graphs are often endowed with probabilities based on relative vertex degree.
A distinction is made between in-degree or out-degree for directed graphs, but this remains silent on ordering.
Nonetheless, a question that arises is whether we can construct a leaf distribution with desired attributes based on 
placing probabilities on vertices  rather than attachments or paths.
We turn to this question next. 

\section{Degree-based  Branching}
\label{sec:branch}

First consider assigning some weights  $\{w\}$ to  vertices. 
Then, starting at the root, traverse the tree by  branching from any vertex to directly  reachable  vertices 
using the relative weights  of the latter as branching probabilities.
The products of such successive branching probabilities induce  a leaf distribution
(equivalently, the destination proportions of a large number of  probabilistically weighted random walks from the root to the leaves).

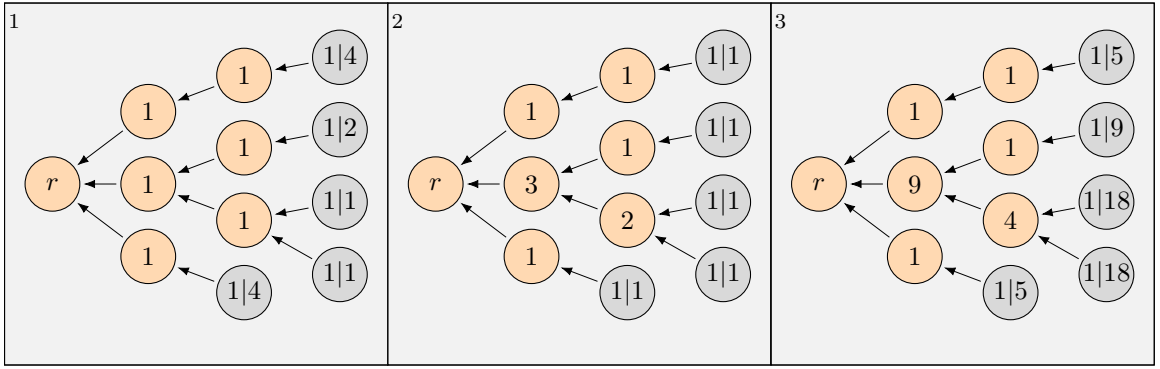
\begin{figure}[tbh]
\centering
\begin{tikzpicture}[xscale=0.6, yscale=0.6,
Myfill/.style ={fill=#1!30, draw=black},
Myfill/.default=orange]

\newcommand{\drawEdge}[2]
{
\draw[Myarrow] let \p1 =  ($(#2) - (#1)$), \n1 = {atan2(\y1,\x1)}, \p2 = ($(\n1:\offset)$) in  ($(#1)+(\p2)$) -- ($(#2)-(\p2)$)
}
\newcommand*{\nx}{8}%
\newcommand*{\ny}{8}%
\newcommand*{\xo}{0}%
\newcommand*{\yo}{\ny/2}%
\newcommand*{\h}{0.1}%
\newcommand*{\xstep}{2.1}%
\newcommand*{\ystep}{1.0}%
\newcommand*{\radius}{0.6}%
\newcommand*{\radiusa}{0.5}%
\newcommand*{\offset}{\radius+\h}%
\newcommand*{\size}{\footnotesize}%
\newcommand*{\sizea}{\footnotesize}%
\newcommand*{\sizeb}{\scriptsize}%
\newcommand*{\tipcol}{gray}%
\newcommand*{\newcol}{white}%
\newcommand*{\backhue}{10}%


\newcommand*{\num}{4}
\newcommand*{\hya}{\ny/5}%
\newcommand*{\hyb}{\ny/5}%
\newcommand*{\hyc}{\ny/5}%

\coordinate (root) at (\xo,\yo);
\coordinate (x1) at ($(root)+(\xstep,-1.0*\hya*\ystep)$);
\coordinate (x2) at ($(x1)+(0,\hya*\ystep)$);
\coordinate (x3) at ($(x2)+(0,\hya*\ystep)$);

\coordinate (y1) at ($(root)+(2*\xstep,-1.5*\hyb*\ystep)$);
\coordinate (y2) at ($(y1)+(0,\hyb*\ystep)$);
\coordinate (y3) at ($(y2)+(0,\hyb*\ystep)$);
\coordinate (y4) at ($(y3)+(0,\hyb*\ystep)$);

\coordinate (z1) at ($(root)+(3*\xstep,-1.25*\hyc*\ystep)$);
\coordinate (z2) at ($(z1)+(0,\hyc*\ystep)$);
\coordinate (z3) at ($(z2)+(0,\hyc*\ystep)$);
\coordinate (z4) at ($(z3)+(0,\hyc*\ystep)$);

\draw [fill=gray!\backhue]  ($(root)-(0.5*\xstep,\yo)$) rectangle +($(\num*\xstep, \ny)$);
\draw [draw=black]  ($(root)-(0.5*\xstep,\yo)$) rectangle +($(\num*\xstep, \ny)$);
\draw ($(root)+(-\xstep/2.5,0.9*\yo)$) node[font=\scriptsize]{1};

\filldraw [Myfill] (root) circle (\radius) node[font=\size]{$r$};
\filldraw [Myfill] (x1) circle (\radius) node[font=\size]{$1$};
\filldraw [Myfill] (x2) circle (\radius) node[font=\size]{$1$};
\filldraw [Myfill] (x3) circle (\radius) node[font=\size]{$1$};

\filldraw [Myfill=\tipcol] (y1) circle (\radius) node[font=\sizea]{$1|4$}; 
\filldraw [Myfill] (y2) circle (\radius) node[font=\size]{$1$}; 
\filldraw [Myfill] (y3) circle (\radius) node[font=\size]{$1$}; 
\filldraw [Myfill] (y4) circle (\radius) node[font=\size]{$1$}; 

\filldraw [Myfill=\tipcol] (z1) circle (\radius) node[font=\sizea]{$1|1$}; 
\filldraw [Myfill=\tipcol] (z2) circle (\radius) node[font=\sizea]{$1|1$}; 
\filldraw [Myfill=\tipcol] (z3) circle (\radius) node[font=\sizea]{$1|2$}; 
\filldraw [Myfill=\tipcol] (z4) circle (\radius) node[font=\sizea]{$1|4$}; 

\drawEdge{x1}{root}; 
\drawEdge{x2}{root}; 
\drawEdge{x3}{root}; 
\drawEdge{y1}{x1}; 
\drawEdge{y2}{x2}; 
\drawEdge{y3}{x2}; 
\drawEdge{y4}{x3}; 
\drawEdge{z1}{y2}; 
\drawEdge{z2}{y2}; 
\drawEdge{z3}{y3}; 
\drawEdge{z4}{y4}; 

\coordinate (root) at (\num*\xstep,\yo);
\coordinate (x1) at ($(root)+(\xstep,-1.0*\hya*\ystep)$);
\coordinate (x2) at ($(x1)+(0,\hya*\ystep)$);
\coordinate (x3) at ($(x2)+(0,\hya*\ystep)$);

\coordinate (y1) at ($(root)+(2*\xstep,-1.5*\hyb*\ystep)$);
\coordinate (y2) at ($(y1)+(0,\hyb*\ystep)$);
\coordinate (y3) at ($(y2)+(0,\hyb*\ystep)$);
\coordinate (y4) at ($(y3)+(0,\hyb*\ystep)$);

\coordinate (z1) at ($(root)+(3*\xstep,-1.25*\hyc*\ystep)$);
\coordinate (z2) at ($(z1)+(0,\hyc*\ystep)$);
\coordinate (z3) at ($(z2)+(0,\hyc*\ystep)$);
\coordinate (z4) at ($(z3)+(0,\hyc*\ystep)$);

\draw [fill=gray!\backhue]  ($(root)-(0.5*\xstep,\yo)$) rectangle +($(\num*\xstep, \ny)$);
\draw [draw=black]  ($(root)-(0.5*\xstep,\yo)$) rectangle +($(\num*\xstep, \ny)$);
\draw ($(root)+(-\xstep/2.5,0.9*\yo)$) node[font=\scriptsize]{2};

\filldraw [Myfill] (root) circle (\radius) node[font=\size]{$r$};
\filldraw [Myfill] (x1) circle (\radius) node[font=\size]{$1$};
\filldraw [Myfill] (x2) circle (\radius) node[font=\size]{$3$};
\filldraw [Myfill] (x3) circle (\radius) node[font=\size]{$1$};

\filldraw [Myfill=\tipcol] (y1) circle (\radius) node[font=\sizea]{$1|1$}; 
\filldraw [Myfill] (y2) circle (\radius) node[font=\size]{$2$}; 
\filldraw [Myfill] (y3) circle (\radius) node[font=\size]{$1$}; 
\filldraw [Myfill] (y4) circle (\radius) node[font=\size]{$1$}; 

\filldraw [Myfill=\tipcol] (z1) circle (\radius) node[font=\sizea]{$1|1$}; 
\filldraw [Myfill=\tipcol] (z2) circle (\radius) node[font=\sizea]{$1|1$}; 
\filldraw [Myfill=\tipcol] (z3) circle (\radius) node[font=\sizea]{$1|1$}; 
\filldraw [Myfill=\tipcol] (z4) circle (\radius) node[font=\sizea]{$1|1$}; 

\drawEdge{x1}{root}; 
\drawEdge{x2}{root}; 
\drawEdge{x3}{root}; 
\drawEdge{y1}{x1}; 
\drawEdge{y2}{x2}; 
\drawEdge{y3}{x2}; 
\drawEdge{y4}{x3}; 
\drawEdge{z1}{y2}; 
\drawEdge{z2}{y2}; 
\drawEdge{z3}{y3}; 
\drawEdge{z4}{y4}; 

\coordinate (root) at (2*\num*\xstep,\yo);
\coordinate (x1) at ($(root)+(\xstep,-1.0*\hya*\ystep)$);
\coordinate (x2) at ($(x1)+(0,\hya*\ystep)$);
\coordinate (x3) at ($(x2)+(0,\hya*\ystep)$);

\coordinate (y1) at ($(root)+(2*\xstep,-1.5*\hyb*\ystep)$);
\coordinate (y2) at ($(y1)+(0,\hyb*\ystep)$);
\coordinate (y3) at ($(y2)+(0,\hyb*\ystep)$);
\coordinate (y4) at ($(y3)+(0,\hyb*\ystep)$);

\coordinate (z1) at ($(root)+(3*\xstep,-1.25*\hyc*\ystep)$);
\coordinate (z2) at ($(z1)+(0,\hyc*\ystep)$);
\coordinate (z3) at ($(z2)+(0,\hyc*\ystep)$);
\coordinate (z4) at ($(z3)+(0,\hyc*\ystep)$);

\draw [fill=gray!\backhue]  ($(root)-(0.5*\xstep,\yo)$) rectangle +($(\num*\xstep, \ny)$);
\draw [draw=black]  ($(root)-(0.5*\xstep,\yo)$) rectangle +($(\num*\xstep, \ny)$);
\draw ($(root)+(-\xstep/2.5,0.9*\yo)$) node[font=\scriptsize]{3};

\filldraw [Myfill] (root) circle (\radius) node[font=\size]{$r$};
\filldraw [Myfill] (x1) circle (\radius) node[font=\size]{$1$};
\filldraw [Myfill] (x2) circle (\radius) node[font=\size]{$9$};
\filldraw [Myfill] (x3) circle (\radius) node[font=\size]{$1$};

\filldraw [Myfill=\tipcol] (y1) circle (\radius) node[font=\sizea]{$1|5$}; 
\filldraw [Myfill] (y2) circle (\radius) node[font=\size]{$4$}; 
\filldraw [Myfill] (y3) circle (\radius) node[font=\size]{$1$}; 
\filldraw [Myfill] (y4) circle (\radius) node[font=\size]{$1$}; 

\filldraw [Myfill=\tipcol] (z1) circle (\radius) node[font=\sizea]{$1|18$}; 
\filldraw [Myfill=\tipcol] (z2) circle (\radius) node[font=\sizea]{$1|18$}; 
\filldraw [Myfill=\tipcol] (z3) circle (\radius) node[font=\sizea]{$1|9$}; 
\filldraw [Myfill=\tipcol] (z4) circle (\radius) node[font=\sizea]{$1|5$}; 

\drawEdge{x1}{root}; 
\drawEdge{x2}{root}; 
\drawEdge{x3}{root}; 
\drawEdge{y1}{x1}; 
\drawEdge{y2}{x2}; 
\drawEdge{y3}{x2}; 
\drawEdge{y4}{x3}; 
\drawEdge{z1}{y2}; 
\drawEdge{z2}{y2}; 
\drawEdge{z3}{y3}; 
\drawEdge{z4}{y4}; 

\end{tikzpicture}
\caption{Example considered earlier with unit weights on leaves shown as the left leaf label and  
1) unit weights on deep vertices
2) each vertex weighted by  the in-degree sum of all vertices that directly attach to it
3) squares of weights in frame 2).
The right leaf label is the leaf distribution induced by branching from root to leaf based on relative vertex weights 
as branching probabilities.} 
\label{fig:branch1}
\end{figure}

In the simplest case, every vertex is taken to have the same unit weight. 
Frame~1 of Figure~\ref{fig:branch1} shows the example considered earlier, 
with unit weights on deep vertices and the left label on the leaves.
The right label on leaves is the unnormalised leaf probability induced by branching from the root as described,
{\it i.e.}\ taking the product of successive branching probabilities.
The leaves at the ends of  isolated chains are allocated the highest probability, contrary to intended behaviour. 

The potential remedy is to allocate higher  weight to vertices exhibiting more branching.
As in the previous section, we need first to traverse the tree backward from the leaves to construct weights.
The difference is that we now assign branch weights to vertices rather than multiplicity weights to attachments.
This is not a likelihood construction exercise since we no longer  retain conditioning on leaves as we traverse the tree backward,
assigning vertex weights.
Hence, once we reach the root, we need to to traverse the tree again from root to leaf  to construct the leaf distribution.

Frame~2 of Figure~\ref{fig:branch1} shows each vertex labelled by weighted in-degree, 
{\it i.e.} the in-degree sum of all vertices that directly attach to it (having assigned  starting values of 1 to all leaves).
In effect, this takes the weights we associated with attachments earlier and places them on the source vertex.
This time the branching distribution at the leaves is also uniform, we have an exact inversion process.

At this point we can try to enforce the qualitative differentiation we seek through further sharpening of relative branch weights.
For example, we might sharpen by raising the weights to some power $\{w^{\alpha}\},~\alpha>1$.
Frame~3 of Figure~\ref{fig:branch1} shows the case $\alpha=2$, {\it i.e.}\ squaring every weight  in Frame~2.
We finally attain a leaf distribution  that is keeping with design objectives.
Notably, isolated chains are assigned the same probability regardless of length.
However, this has come at the expense of imposing non-linear structure on linear weighted in-degree 
(like Krapivsky {\it et al.}~\cite{Krapivsky}).

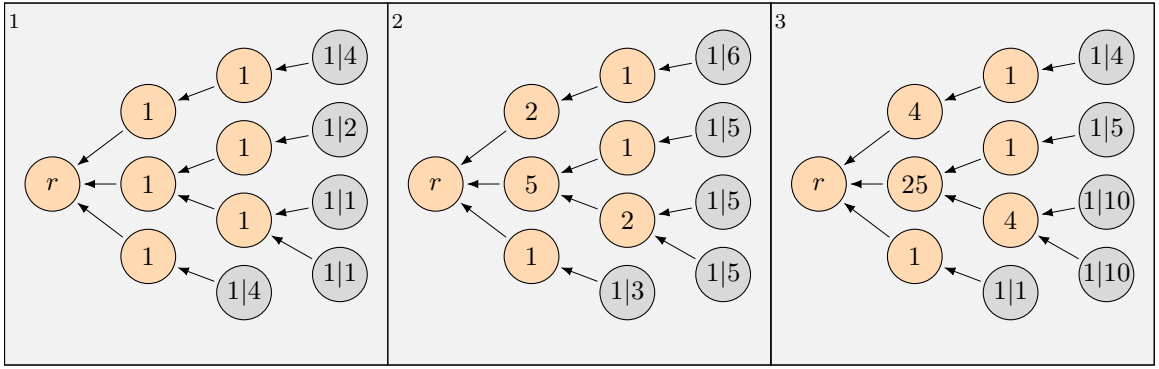
\begin{figure}[tbh]
\centering
\begin{tikzpicture}[xscale=0.6, yscale=0.6,
Myfill/.style ={fill=#1!30, draw=black},
Myfill/.default=orange]

\newcommand{\drawEdge}[2]
{
\draw[Myarrow] let \p1 =  ($(#2) - (#1)$), \n1 = {atan2(\y1,\x1)}, \p2 = ($(\n1:\offset)$) in  ($(#1)+(\p2)$) -- ($(#2)-(\p2)$)
}
\newcommand*{\nx}{8}%
\newcommand*{\ny}{8}%
\newcommand*{\xo}{0}%
\newcommand*{\yo}{\ny/2}%
\newcommand*{\h}{0.1}%
\newcommand*{\xstep}{2.1}%
\newcommand*{\ystep}{1.0}%
\newcommand*{\radius}{0.6}%
\newcommand*{\radiusa}{0.5}%
\newcommand*{\offset}{\radius+\h}%
\newcommand*{\size}{\footnotesize}%
\newcommand*{\sizea}{\footnotesize}%
\newcommand*{\sizeb}{\scriptsize}%
\newcommand*{\tipcol}{gray}%
\newcommand*{\newcol}{white}%
\newcommand*{\backhue}{10}%


\newcommand*{\num}{4}
\newcommand*{\hya}{\ny/5}%
\newcommand*{\hyb}{\ny/5}%
\newcommand*{\hyc}{\ny/5}%

\coordinate (root) at (\xo,\yo);
\coordinate (x1) at ($(root)+(\xstep,-1.0*\hya*\ystep)$);
\coordinate (x2) at ($(x1)+(0,\hya*\ystep)$);
\coordinate (x3) at ($(x2)+(0,\hya*\ystep)$);

\coordinate (y1) at ($(root)+(2*\xstep,-1.5*\hyb*\ystep)$);
\coordinate (y2) at ($(y1)+(0,\hyb*\ystep)$);
\coordinate (y3) at ($(y2)+(0,\hyb*\ystep)$);
\coordinate (y4) at ($(y3)+(0,\hyb*\ystep)$);

\coordinate (z1) at ($(root)+(3*\xstep,-1.25*\hyc*\ystep)$);
\coordinate (z2) at ($(z1)+(0,\hyc*\ystep)$);
\coordinate (z3) at ($(z2)+(0,\hyc*\ystep)$);
\coordinate (z4) at ($(z3)+(0,\hyc*\ystep)$);

\draw [fill=gray!\backhue]  ($(root)-(0.5*\xstep,\yo)$) rectangle +($(\num*\xstep, \ny)$);
\draw [draw=black]  ($(root)-(0.5*\xstep,\yo)$) rectangle +($(\num*\xstep, \ny)$);
\draw ($(root)+(-\xstep/2.5,0.9*\yo)$) node[font=\scriptsize]{1};

\filldraw [Myfill] (root) circle (\radius) node[font=\size]{$r$};
\filldraw [Myfill] (x1) circle (\radius) node[font=\size]{$1$};
\filldraw [Myfill] (x2) circle (\radius) node[font=\size]{$1$};
\filldraw [Myfill] (x3) circle (\radius) node[font=\size]{$1$};

\filldraw [Myfill=\tipcol] (y1) circle (\radius) node[font=\sizea]{$1|4$}; 
\filldraw [Myfill] (y2) circle (\radius) node[font=\size]{$1$}; 
\filldraw [Myfill] (y3) circle (\radius) node[font=\size]{$1$}; 
\filldraw [Myfill] (y4) circle (\radius) node[font=\size]{$1$}; 

\filldraw [Myfill=\tipcol] (z1) circle (\radius) node[font=\sizea]{$1|1$}; 
\filldraw [Myfill=\tipcol] (z2) circle (\radius) node[font=\sizea]{$1|1$}; 
\filldraw [Myfill=\tipcol] (z3) circle (\radius) node[font=\sizea]{$1|2$}; 
\filldraw [Myfill=\tipcol] (z4) circle (\radius) node[font=\sizea]{$1|4$}; 

\drawEdge{x1}{root}; 
\drawEdge{x2}{root}; 
\drawEdge{x3}{root}; 
\drawEdge{y1}{x1}; 
\drawEdge{y2}{x2}; 
\drawEdge{y3}{x2}; 
\drawEdge{y4}{x3}; 
\drawEdge{z1}{y2}; 
\drawEdge{z2}{y2}; 
\drawEdge{z3}{y3}; 
\drawEdge{z4}{y4}; 

\coordinate (root) at (\num*\xstep,\yo);
\coordinate (x1) at ($(root)+(\xstep,-1.0*\hya*\ystep)$);
\coordinate (x2) at ($(x1)+(0,\hya*\ystep)$);
\coordinate (x3) at ($(x2)+(0,\hya*\ystep)$);

\coordinate (y1) at ($(root)+(2*\xstep,-1.5*\hyb*\ystep)$);
\coordinate (y2) at ($(y1)+(0,\hyb*\ystep)$);
\coordinate (y3) at ($(y2)+(0,\hyb*\ystep)$);
\coordinate (y4) at ($(y3)+(0,\hyb*\ystep)$);

\coordinate (z1) at ($(root)+(3*\xstep,-1.25*\hyc*\ystep)$);
\coordinate (z2) at ($(z1)+(0,\hyc*\ystep)$);
\coordinate (z3) at ($(z2)+(0,\hyc*\ystep)$);
\coordinate (z4) at ($(z3)+(0,\hyc*\ystep)$);

\draw [fill=gray!\backhue]  ($(root)-(0.5*\xstep,\yo)$) rectangle +($(\num*\xstep, \ny)$);
\draw [draw=black]  ($(root)-(0.5*\xstep,\yo)$) rectangle +($(\num*\xstep, \ny)$);
\draw ($(root)+(-\xstep/2.5,0.9*\yo)$) node[font=\scriptsize]{2};

\filldraw [Myfill] (root) circle (\radius) node[font=\size]{$r$};
\filldraw [Myfill] (x1) circle (\radius) node[font=\size]{$1$};
\filldraw [Myfill] (x2) circle (\radius) node[font=\size]{$5$};
\filldraw [Myfill] (x3) circle (\radius) node[font=\size]{$2$};

\filldraw [Myfill=\tipcol] (y1) circle (\radius) node[font=\sizea]{$1|3$}; 
\filldraw [Myfill] (y2) circle (\radius) node[font=\size]{$2$}; 
\filldraw [Myfill] (y3) circle (\radius) node[font=\size]{$1$}; 
\filldraw [Myfill] (y4) circle (\radius) node[font=\size]{$1$}; 

\filldraw [Myfill=\tipcol] (z1) circle (\radius) node[font=\sizea]{$1|5$}; 
\filldraw [Myfill=\tipcol] (z2) circle (\radius) node[font=\sizea]{$1|5$}; 
\filldraw [Myfill=\tipcol] (z3) circle (\radius) node[font=\sizea]{$1|5$}; 
\filldraw [Myfill=\tipcol] (z4) circle (\radius) node[font=\sizea]{$1|6$}; 

\drawEdge{x1}{root}; 
\drawEdge{x2}{root}; 
\drawEdge{x3}{root}; 
\drawEdge{y1}{x1}; 
\drawEdge{y2}{x2}; 
\drawEdge{y3}{x2}; 
\drawEdge{y4}{x3}; 
\drawEdge{z1}{y2}; 
\drawEdge{z2}{y2}; 
\drawEdge{z3}{y3}; 
\drawEdge{z4}{y4}; 

\coordinate (root) at (2*\num*\xstep,\yo);
\coordinate (x1) at ($(root)+(\xstep,-1.0*\hya*\ystep)$);
\coordinate (x2) at ($(x1)+(0,\hya*\ystep)$);
\coordinate (x3) at ($(x2)+(0,\hya*\ystep)$);

\coordinate (y1) at ($(root)+(2*\xstep,-1.5*\hyb*\ystep)$);
\coordinate (y2) at ($(y1)+(0,\hyb*\ystep)$);
\coordinate (y3) at ($(y2)+(0,\hyb*\ystep)$);
\coordinate (y4) at ($(y3)+(0,\hyb*\ystep)$);

\coordinate (z1) at ($(root)+(3*\xstep,-1.25*\hyc*\ystep)$);
\coordinate (z2) at ($(z1)+(0,\hyc*\ystep)$);
\coordinate (z3) at ($(z2)+(0,\hyc*\ystep)$);
\coordinate (z4) at ($(z3)+(0,\hyc*\ystep)$);

\draw [fill=gray!\backhue]  ($(root)-(0.5*\xstep,\yo)$) rectangle +($(\num*\xstep, \ny)$);
\draw [draw=black]  ($(root)-(0.5*\xstep,\yo)$) rectangle +($(\num*\xstep, \ny)$);
\draw ($(root)+(-\xstep/2.5,0.9*\yo)$) node[font=\scriptsize]{3};

\filldraw [Myfill] (root) circle (\radius) node[font=\size]{$r$};
\filldraw [Myfill] (x1) circle (\radius) node[font=\size]{$1$};
\filldraw [Myfill] (x2) circle (\radius) node[font=\size]{$25$};
\filldraw [Myfill] (x3) circle (\radius) node[font=\size]{$4$};

\filldraw [Myfill=\tipcol] (y1) circle (\radius) node[font=\sizea]{$1|1$}; 
\filldraw [Myfill] (y2) circle (\radius) node[font=\size]{$4$}; 
\filldraw [Myfill] (y3) circle (\radius) node[font=\size]{$1$}; 
\filldraw [Myfill] (y4) circle (\radius) node[font=\size]{$1$}; 

\filldraw [Myfill=\tipcol] (z1) circle (\radius) node[font=\sizea]{$1|10$}; 
\filldraw [Myfill=\tipcol] (z2) circle (\radius) node[font=\sizea]{$1|10$}; 
\filldraw [Myfill=\tipcol] (z3) circle (\radius) node[font=\sizea]{$1|5$}; 
\filldraw [Myfill=\tipcol] (z4) circle (\radius) node[font=\sizea]{$1|4$}; 

\drawEdge{x1}{root}; 
\drawEdge{x2}{root}; 
\drawEdge{x3}{root}; 
\drawEdge{y1}{x1}; 
\drawEdge{y2}{x2}; 
\drawEdge{y3}{x2}; 
\drawEdge{y4}{x3}; 
\drawEdge{z1}{y2}; 
\drawEdge{z2}{y2}; 
\drawEdge{z3}{y3}; 
\drawEdge{z4}{y4}; 

\end{tikzpicture}
\caption{Same example with unit weights on leaves shown as the left leaf label and  
1) unit weights on deep vertices
2) each vertex weighted by by cumulative in-degree from the leaves
3) squares of weights in frame 2).
The right leaf label is the leaf distribution induced by branching from root to leaf based on relative vertex weights 
as branching probabilities.} 
\label{fig:branch2}
\end{figure}

We consider next using the weight of the subtree rooted at each vertex,
as referred to earlier when we discussed the blockchain method due to  Sompolinsky and Zohar~\cite{SompZohar1}.
We define it  as the total number of vertices  that the branch carries,
{\it i.e.}\ the   cumulative in-degree of the vertex as a root of the subtree 
(we could choose to include the latter vertex  so that  weight = cumulative in-degree$+1$).

Frame~2 of Figure~\ref{fig:branch2} shows vertices weighted by cumulative in-degree.
The leaf terminating the longer isolated vertex still receives highest probability.
Despite having lower branch weight, it does not have to allocate that weight to subsidiary branches, resulting  in the single leaf 
inheriting all of it.
Once again,  further sharpening of relative branch weights is indicated.
Frame~3 of Figure~\ref{fig:branch2} shows the case $\alpha=2$, {\it i.e.}\ squaring every subtree weight  in Frame~2.
Only then  do we finally attain a leaf distribution  that is keeping with design objectives,
although in this case  longer isolated chains are preferred over shorter ones.
Such  `nonlinear sharpening' of cumulative in-degree weights is the  approach described by Popov~\cite{Popov1}, 
where the subtree weights are exponentiated, $\{\exp(\alpha w)\}$.

In both examples, we may think of the 3 successive frames as corresponding to exponents $\{w^{\alpha}\},~\alpha =0,1,2$ respectively
for the particular choice of vertex weights. 
In our view, introducing  nonlinear functions of  weighted degree or cumulative degree 
in order  to meet design objectives is rather {\it ad hoc}.
We have included this section for conceptual completeness but we shall not pursue degree-based branching further here.

\section{Computational Examples}
\label{sec:comp}
The growth of random trees by leaf attachment is best illustrated by visual animation.
Here, we can only display 
still  frames from animations available on the site
\href{https://drive.google.com/drive/folders/1ENnABnUrhUGC8eq4YeT1Vmedce6pZiWu}{Sibisi~Movies}.
We wrote the  tree generation code in C$++$  and created the animations  using  {\rm \sffamily gnuplot}~\cite{gnuplot}.
Although the code is quite general, the examples here are of modest scale to aid visual interpretation.
The runtime parameters are: 
\begin{labeling}{Bayes}
\setlength{\topsep}{-4pt}
\setlength{\itemsep}{0pt}
\setlength{\parskip}{0pt}
\setlength{\parsep}{0pt} 
\item[$N$]  The number of time intervals $\{t=0,1\ldots N-1\}$
\item[$\mu$] The  mean of the Poisson distribution used to generate the number of new vertices in each time interval $t>0$.
In the examples here, $\mu$ is held constant for all $t>0$. 
The initial state at $t=0$ is the root as the sole  new vertex.
\item[Bayes] The choice of leaf distribution used to sample attachment points to grow the tree: 
\begin{labeling}{\,Case 0:}
\setlength{\topsep}{-4pt}
\setlength{\itemsep}{0pt}
\setlength{\parskip}{0pt}
\setlength{\parsep}{0pt} 
\item[\,Case 0:]  The  prior $\Pr(\ell)$, taken to be uniform
\item[\,Case 1:]  The  posterior $\Pr(\ell|\cH)$, using the `global' likelihood of subsection~\ref{sec:global}
\item[\,Case 2:]  The  posterior $\Pr(\ell|\cH)$, using the `local' likelihood of subsection~\ref{sec:local}
\end{labeling}
\end{labeling}

\subsection{First Example}
\label{example1}
\begin{figure}[tbh]
\centering
\input{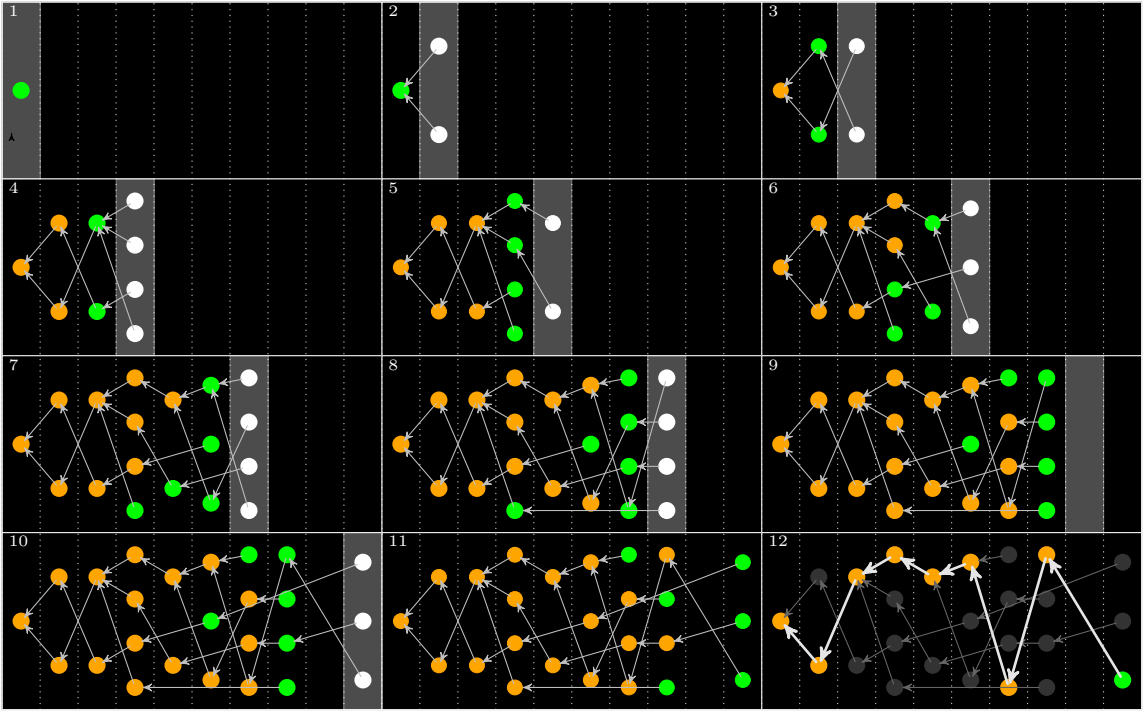}
    \vspace*{-3mm}
\caption{Growth using uniform leaf distribution. Poisson mean 2, 10 time intervals.
Final frame shows longest path.} 
\label{fig:plot10_0}
\end{figure}
Choose $\mu=2, N=10$ and use a fixed  seed to generate the same Poisson  sequence of new vertices 
for each of the 3  {\rm \sffamily Bayes} cases.

\begin{figure}[tbh]
\centering
\input{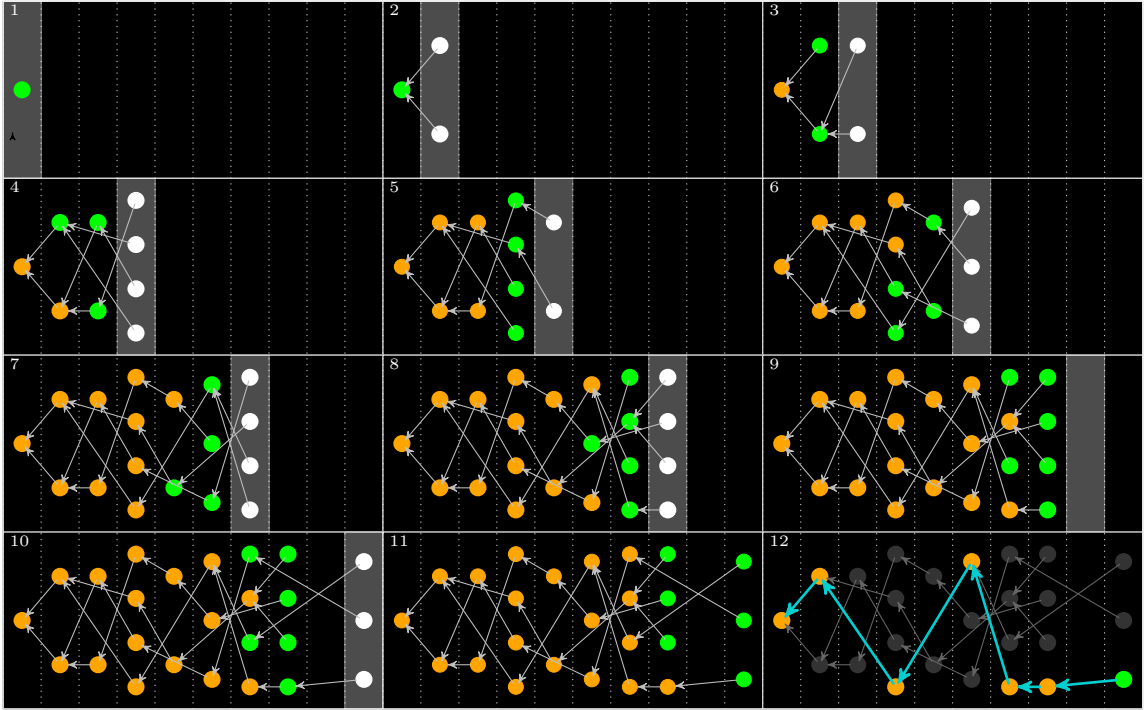}
    \vspace*{-3mm}
\caption{Growth using posterior leaf distribution with global likelihood. Poisson mean 2, 10 time intervals. 
Final frame shows path  terminated by leaf with highest posterior}
\label{fig:plot10_1}
\end{figure}
\begin{figure}[tbh]
\centering
\input{plot10_2}
    \vspace*{-3mm}
\caption{Growth using posterior leaf distribution with local likelihood. Poisson mean 2, 10 time intervals.
Final frame shows path  terminated by leaf with highest posterior.}
\label{fig:plot10_2}
\end{figure}

Figure~\ref{fig:plot10_0} shows consecutive movie frames 1 to 10 
as the tree grows from the root by uniform leaf attachment ({\rm \sffamily Bayes}=0).
Frame~11 shows the complete tree after the last set of new vertices have been converted to leaves.
Frame~12 shows the longest path  between leaf and root (where this is not unique, we simply choose one). 
The intention is not to abandon the rest of the tree but to find a  proxy for the connectivity  structure of the tree that is easy to visualise.
This is a particularly helpful attribute for trees much denser than the sparse, low scale (small $\mu, N$) 
example chosen here for illustrative purposes.

Figures~\ref{fig:plot10_1} and \ref{fig:plot10_2} show the analogous growth for 
{\rm \sffamily Bayes}=1 and {\rm \sffamily Bayes}=2 respectively.
For these two cases, the final frame shows the path to the root starting at  the leaf of maximum posterior probability
(if this is not unique, we display one such path).
In this case of a uniform prior, the maximum posterior path coincides with the maximum likelihood path, 
equivalently, the heaviest path.
We can already visually detect the preference in Figure~\ref{fig:plot10_2} for attachment to well-connected leaves
and tending to leave  isolated chains behind.

\subsection{Second Example}
\label{example1}
Chooose $\mu=2, N=25$ and use a fixed  seed to generate the same Poisson  sequence of new vertices 
for each of the three  {\rm \sffamily Bayes} cases.

\begin{figure}[tbh]
\centering
\input{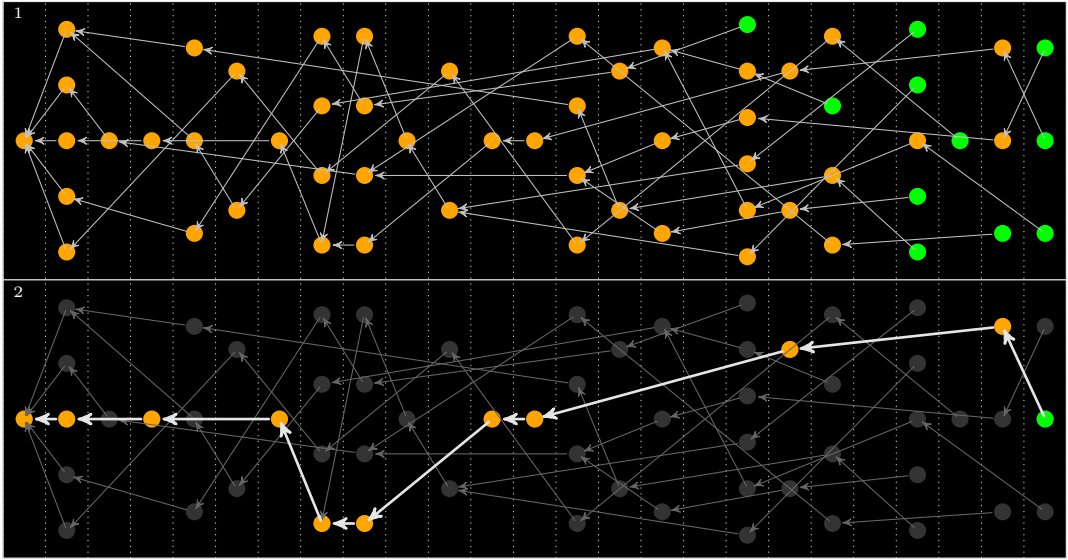}
\caption{Growth using uniform leaf distribution. Poisson mean 2, 25 time intervals. 
(1) Last frame in sequence. (2) Frame showing longest path.} 
\label{fig:plot25_0}
\end{figure}

Figure~\ref{fig:plot25_0}  shows the final frame of the 25-frame growth sequence for the {\rm \sffamily Bayes}=0 case.
Since the leaf attachment distribution is uniform and  independently of the historical connectivity of the tree,
one can expect attachments of any length to leaves of any connectivity or age. 
Hence the longest path unsurprisingly contains a  mix of short and long attachments and is thus rather sparse (10 attachments) 
compared to the time sequence of 25 intervals.

\begin{figure}[tbh]
\centering
\input{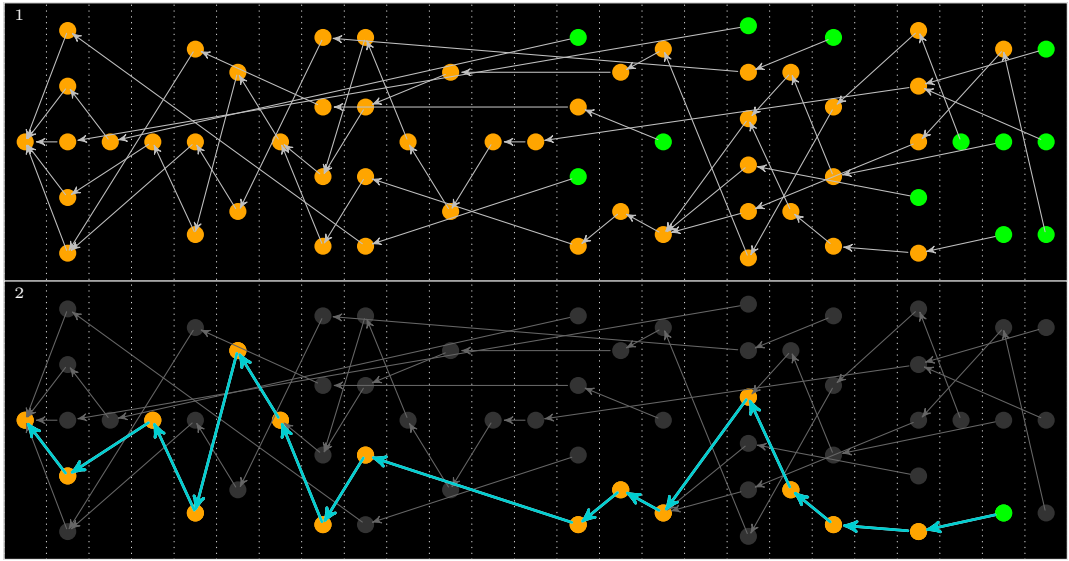}
\caption{Growth using posterior leaf distribution with global likelihood. Poisson mean 2, 25 time intervals. 
(1) Last frame in sequence. (2) Frame showing path of highest posterior.} 
\label{fig:plot25_1}
\end{figure}

Figure~\ref{fig:plot25_1}  shows the final frame of the sequence for the {\rm \sffamily Bayes}=1 case.
The maximum posterior path is  denser (15 attachments) than the longest path of the  {\rm \sffamily Bayes}=0 case,
in keeping with the preference for attaching to  better connected leaves as the tree grows.

\begin{figure}[tbh]
\centering
\input{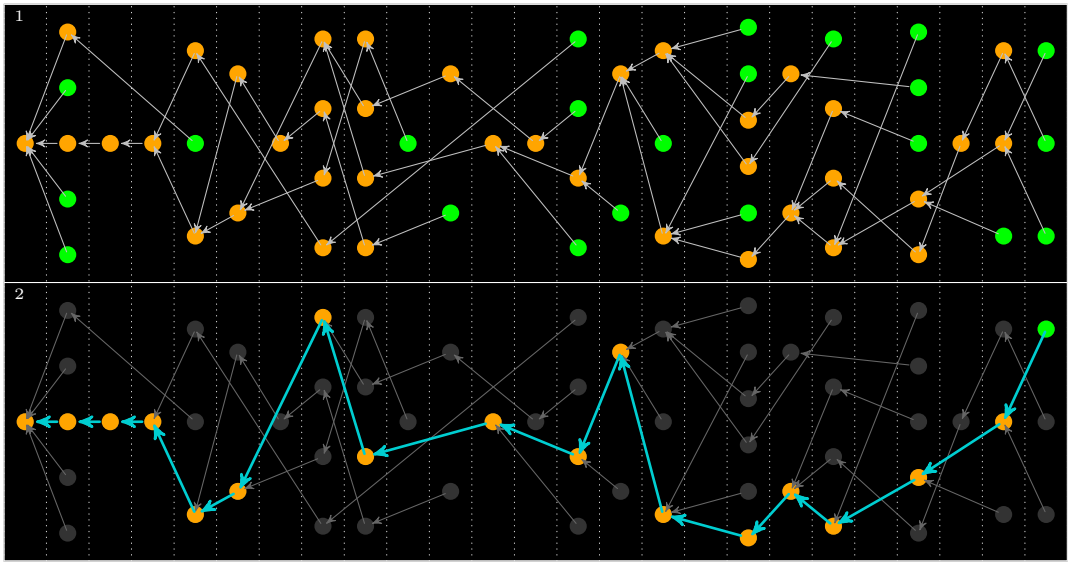}
\caption{Growth using posterior leaf distribution with local likelihood. Poisson mean 2, 25 time intervals. 
Note absence of long attachments to old leaves.
(1) Last frame in sequence. (2) Frame showing path of highest posterior.} 
\label{fig:plot25_2}
\end{figure}

Figure~\ref{fig:plot25_2}  shows the final frame of the sequence for the {\rm \sffamily Bayes}=2 case.
There are now hardly any long attachments and the maximum posterior path is  even denser (17 attachments)  
than the  {\rm \sffamily Bayes}=1 case.

\subsection{Discussion}
\label{sec:discuss}
As the examples illustrate, sampling from a uniform prior leaf distribution $\Pr(\ell)$ allows long attachments to old leaves, 
which enjoy the same probability of attachment as any other leaf.
By contrast, the posterior leaf distribution $\Pr(\ell|\cH)$ tends to discourage long attachments to old vertices 
(particularly for  {\rm \sffamily Bayes}=2) and encourage short attachments.
 This leads to  dense, well-connected  paths at the expense of isolated  paths that are left behind 
 (but not definitively abandoned because all leaves have non-zero probability).

The caveat is that we have only presented a few animations to indicate the behaviour of the different growth schemes.
Further exploration would need to be guided by specific questions.
For example, we might wish to quantify the attachment length of the longest  path ({\rm \sffamily Bayes}=1) or highest probability  path
({\rm \sffamily Bayes}=1,2), as an average over an ensemble of animations at given $\mu, N$. 
We might then study the evolution  of such an average path length with increasing $N$ and compare the efficiency of the different attachment schemes at creating dense paths.

 A compromise between growth  properties induced by $\Pr(\ell)$  and $\Pr(\ell|\cH)$, should we deem it necessary,
 might be to use a linear combination of the two
 \begin{equation}
 q \Pr(\ell)+ (1-q) \Pr(\ell|\cH) \qquad {\rm for}\quad  0\le q\le 1
 \label{eq:binarymix}
 \end{equation}
We might even imagine a time-dependent $q(t)$,  alternately favouring    
$\Pr(\ell|\cH)$ to encourage attachment to well-connected leaves and 
$\Pr(\ell)$ to give all leaves equal chance of receiving attachments.
We might think of a  smoothly  oscillating $q(t)$ as a source of  `breathing' growth, 
where~(\ref{eq:binarymix}) periodically tightens to favour  $\Pr(\ell|\cH)$
and relaxes to favour  $\Pr(\ell)$.

\section{Conclusion}
\label{sec:conclusion}
We have presented a novel Bayesian method to grow a rooted time-ordered directed  tree
 by probabilistic leaf attachment.
 The   connectivity of the tree induces the likelihood function which, combined with an assigned prior distribution, 
 leads to the posterior leaf distribution from which we draw attachment points.
 
 The concept of a directed time-ordered path has played a fundamental role in defining the  connectivity of the tree,
 which is effectively defined as the merging of such paths from leaves to the root.
The directed time-ordered path will remain the fundamental construct as we generalise from a tree to a graph 
where a new vertex may make  multiple attachments.
There will then be multiple paths from leaf to root, with both merging and splitting of paths in between.

It is also worth exploring the effect of  an incomplete path history from leaves, possibly terminating after a specified 
number of attachments before reaching the root.

A further  question that arises naturally  is whether we can modify the leaf attachment methodology 
in order to attach to vertices at any depth.
Given a deep vertex $v$, we might then consider both \\
a) directed paths from $v$   to the root (as we did for the case where $v$ is a leaf) and   \\
b) reverse directed paths from $v$ to the leaves of the current snapshot of the tree. \\
Both sets of paths would then inform construction of the likelihood.

Such a construction would  equip us to model the citation network discussed in section~\ref{sec:context}. 
We intend to pursue  these ideas further in a dedicated paper.

\clearpage
\subsection*{Acknowledgement}
We wrote this paper  in \LaTeX\ using the graph drawing package PGF/Ti{\it k}\,Z to generate all diagrams.
We thank Sibusiso Sibisi for helpful discussions and input on the C++ code and the use of the {\rm \sffamily gnuplot} graphing package 
for the computational examples.
  
\bibliography{Leaf}{}
\bibliographystyle{plain} 

\end{document}